\documentclass[12pt]{iopart}
\eqnobysec
\usepackage{graphicx}
\usepackage{color}

\newcommand{\be}{\begin{equation}}
\newcommand{\ee}{\end{equation}}

\newcommand{\vecp}{{\mathbf p}}
\newcommand{\vecq}{{\mathbf q}}

\newcommand{\x}{{\mathbf x}}

\newcommand{\X}{{\mathbf X}}

\newcommand{\y}{{\mathbf y}}
\newcommand{\Y}{{\mathbf Y}}

\newcommand{\J}{{\mathbf J}}

\newcommand{\mH}{{\mathbf H}}

\newcommand{\Id}{{\mathbf I}}

\newcommand{\der}{\partial}

\newcommand{\vct}[1]{\ensuremath\mbox{\boldmath$ #1 $}}
\newcommand{\Vxi}{{\vct \xi}}
\newcommand{\Veta}{\vct \eta}

\begin{document}

\title{Energy transitions driven by phase space reflection operators}

\author {Alfredo M. Ozorio de Almeida\footnote{ozorio@cbpf.br}}
\address{Centro Brasileiro de Pesquisas Fisicas,
Rua Xavier Sigaud 150, 22290-180, Rio de Janeiro, R.J., Brazil}

\begin{abstract}

Phase space reflection operators lie at the core of the Wigner-Weyl representation of density operators and observables. The role of the corresponding classical reflections is known in the construction of semiclassical approximations to Wigner functions of pure eigenstates and their 
coarsegrained microcanonical superpositions, which are not restricted to classically integrable systems. In their active role as unitary operators, they generate transitions between pairs of eigenstates specified by transition Wigner functions (or cross-Wigner functions): The square modulus of the transition Wigner function at each point in phase space is the transition probability for the reflection through that point.

Coarsegraining the initial and final energies provides a transition probability density as a phase space path integral. It is here investigated in the simplest classical approximation involving microcanonical Wigner functions. A reflection operator generates a transition between a pair of energy shells with a probability density given by the integral of the inverse modulus of a Poisson bracket over the intersection of a shell with the reflection of its pair. The singularity of the pair of Wigner functions at their dominant caustics is nicely integrable over their intersection, except for a single degree of freedom. Even though this case is not directly relevant for future investigations of chaotic systems, it is shown here how the improved approximation of the spectral Wigner functions in terms of Airy functions resolves the singularity.

\end{abstract}

\maketitle

\section{Introduction}

The origins of semiclassical (SC) approximations in quantum mechanics go back to the old quantum theory, described in terms of action-angle variables of integrable classical systems. Their unitary evolution is well described for a finite time by the extensions of Van Vleck's original theory \cite{VanVleck,Gutzbook}, whatever the nature of the driving Hamiltonian, which may range from integrable, to mixed chaotic, or hard chaotic characterized by general hyperbolic separation between neighbouring trajectories. Usually, the initial states that are prepared in experimental setups are indeed simple, such as coherent states, or integrable though highly excited states of Rydberg atoms, but it is legitimate to enquire about the relation of the evolution of nonintegrable states to their corresponding classical structure. 

The full spectrum of eigenenergies of hard chaotic systems has been the subject of intensive study, beautifully reviewed by Fritz Haake and collaborators \cite{Haake}. It is not readily accessible through finite time SC propagation, demanding resummation over periodic trajectories 
\cite{Bog90, BerKea} and this is also proposed for their individual eigenstates \cite{AgFish}. 
On the other hand, smoothed spectra and coarse-grained microcanonical (mixed) density operators 
\cite{Bogscars,Ber89,Ber89b,Report}, are available, without any restriction to hard chaos. The extended Van Vleck theory does not cope with such initial states, whether coarse-grained or resummed, so we do not have access so far to matrix elements of an external operator, or to the probability of energy transitions that it may produce. 

Percival and Richards \cite{PerRich} obtained, through the Heisenberg correspondence principle, off-diagonal matrix elements of an observable, in systems with a single degree of freedom, as a Fourier integral over the angle for the action fixed halfway between that of both states. This was later extended to general integrable systems \cite{Oz84} by way of the transition Wigner function (or cross-Wigner function) described bellow.

A notable feature of the Wigner-Weyl representation is the role played by the family of phase space reflection operators $\hat{R}_\x$. 
These unitary operators correspond to the classical reflection $\x-\X \mapsto \x+\X$ in the $2N$-dimensional {\it phase space} with coordinates 
$\x = (\vecp,\vecq)=(p_1,...,p_{N},q_1,...,q_{N})$. Since this aspect is not widely appreciated, it is reviewed in Appendix A.
The square of a reflection is the identity, that is,  $\hat{R}_\x^2 = \Id$, the identity operator, so that the eigenvalues of the reflection operators are  $\pm 1$. Thus, ${\hat R}_\x$ are also observables, albeit lacking classical correspondents in this role. The point is that Grossman \cite{Grossmann} and Royer \cite{Royer} showed that, given any density operator $\hat{\rho}$,
\footnote{Wigner-Weyl formulae here follow the notation in \cite{SarOA19}, which differ by factors of $2^N$ from my earlier work, 
such as \cite{Report}.}
\be
W(\x) \equiv \frac{1}{(2\pi\hbar)^N}~ {\rm tr}~\hat{\rho}~\hat{R}_{\x}
\label{Wigfn}
\ee
defines its Wigner function \cite{Wigner} and similarly one obtains the phase space function $O(\x)$ that represents any operator $\hat O$.

The original motivation for Wigner was to express expectations as ordinary classical averages, that is,
\be
\langle \hat{O} \rangle = {\rm tr}~\hat{\rho}~\hat{O} = \int {\rm d}^{2N}\x ~ W (\x)~O(\x),
\label{average}
\ee 
but, for a pure eigenstate $|k\rangle$ of a Hamiltonian $\hat H$ with energy $E_k$, the expectation of a reflection is simply
\be
\langle \hat{R}_{\x} \rangle(k) = \langle k|\hat{R}_{\x}|k\rangle =(2\pi\hbar)^N W_k(\x) ~.
\ee
The possibility of experimentally measuring the Wigner function as an average over the eigenvalues of reflection operators 
\cite{Englert,LutterbachDav} was achieved by Bertet et al \cite{Bertet02}.
In contrast, one can emphasize the alternative unitary role of reflection operators, so as to consider
\be 
P_k(\x)  = |\langle k|\hat{R}_{\x}|k\rangle|^2 = (2\pi\hbar)^{2N}|W_k(\x)|^2
\ee
as the probability of measuring the energy $E_k$ in the reflected state $|k\rangle_{\x} = \hat{R}_{\x}|k\rangle$.

Though less familiar, the Grossman-Royer expression for the {\it transition Wigner function} or {\it cross Wigner function} between a pair of pure states (here selected as
energy eigenstates $|k\rangle$ and $|l\rangle$) is
\be
W_{kl}(\x) \equiv \frac{1}{(2\pi\hbar)^N}~ {\rm tr}~|l\rangle\langle k| ~\hat{R}_{\x} = \frac{1}{(2\pi\hbar)^N}~ \langle k|\hat{R}_{\x}|l\rangle ~.
\label{trWigfn}
\ee
The matrix elements of general operators are expressed as
\be
\langle k| \hat{O}|l \rangle = {\rm tr}~ |l\rangle\langle k|~\hat{O} = \int {\rm d}^{2N}\x ~ W_{kl} (\x)~O(\x) ~.
\label{Melements}
\ee 

In contrast to ordinary Wigner functions, the transition Wigner function for a fixed the pair of states integrates to zero, 
since ${\rm tr}~|l\rangle\langle k|=0$. Alternatively, for a fixed reflection centre $\x$, the concept of $W_{kl}(\x)$ as a matrix of all the energy transition amplitudes $E_k \mapsto E_l$ from $|k\rangle$ to $|l\rangle$, including $W_{kk}(\x)= W(\x)$ as a diagonal element, was called the {\it Moyal matrix} in \cite{Oz84}, 
with the respective probabilities
\be
P_{kl}(\x) = (2\pi\hbar)^{2N} ~ |W_{kl}(\x)|^2 ~.
\ee
The interpretation of the cross Wigner function as a transition amplitude then leads immediately to the sum rule:
\be
\sum_l ~P_{kl}(\x) = (2\pi\hbar)^{2N} ~ \sum_l  ~ |W_{kl}(\x)|^2 =1
\ee
for any $k$ and each fixed reflection centre $\x$.

Direct access to the transition probability from ordinary Wigner functions through the pure state identity
\be
P_{kl}(\x) = (2\pi\hbar)^{N} \int {\rm d}^{2N}\X ~ W_k(\x+\X)~W_l(\x-\X) ~,
\label{Ptrans0}
\ee
has been recently derived
\footnote{The derivation for a discrete phase space in \cite{SarOA19} follows that of a set of identities in the usual continuous phase space presented in \cite{SarOA16}. }.
In the diagonal case this takes the form of the particular identity \cite{SarOA16}
\be
P_{kk}(\x) = (2\pi\hbar)^{N} \int {\rm d}^{2N}\X ~ W_k(\x+\X)~W_k(\x-\X) = W_k(\x)^2 ~.
\label{Pkk}
\ee
It will be our main tool in the construction of approximations of coarsegrained energy transition probability densities.

As it stands for the last few decades, we have standard semiclassical (SC) approximations for $W_k(\x)$ if $N=1$, but their extension
for $N>1$ is limited to integrable classical systems \cite{Ber77,AlmHan82}.
General results that encompass classical chaos were only achieved for the spectral Wigner function \cite{Ber89,Ber89b,Report}
\be
W_E(\x,\epsilon) \equiv (2\pi\hbar)^N \sum_k \delta_\epsilon(E-E_k)~ W_k(\x),
\ee
which represents the spectral density operator 
\be
{\hat \rho}_E(\epsilon) \equiv \sum_k \delta_\epsilon(E-E_k)~ |k\rangle \langle k|,
\ee
for a classically narrow energy range $\epsilon$ centred on the energy E, such that 
\be
\delta_\epsilon(E)\equiv \frac{1}{\pi} ~ \frac{\epsilon}{\epsilon^2 + E^2}
\label{widelta}
\ee 
integrates as a Dirac $\delta$-function. The phase space integral over the spectral Wigner function elicits the smoothed density of states 
\be
\nu(E,\epsilon) \equiv \frac{1}{2\pi\hbar}\int {\rm d}\x~ W_E(\x,\epsilon) = \frac{1}{\pi} \sum_k \frac{\epsilon}{\epsilon^2 + (E-E_k)^2}~.
\ee
Alternatively, for each fixed energy, this can be considered as a coarsegrained microcanonical partition function, 
so that the normalized {\it microcanonical Wigner functions} are $ W_E(\x,\epsilon) /\nu(E,\epsilon) $.

The extension of the SC approximation to the transition Wigner function was presented in \cite{Oz84}, though again only for classically integrable systems. The difficulty in generalizing the recipe for coarsegrained spectral Wigner functions is that no analogous results are available for the direct superposition of transition Wigner functions. Even so, summing through neighbouring pairs of pure states in the identity \eref{Ptrans0} supplies the probability density for a coarse grained transition $E \mapsto E' $ as
\begin{eqnarray}
\fl P_{EE'}(\x, \epsilon)  
\equiv (2\pi\hbar)^{N}\sum_{k,l}  \delta_\epsilon(E-E_k)~\delta_\epsilon(E'-E_l) \int {\rm d}^{2N}\X ~ W_k(\x+\X)~W_l(\x-\X) \\ \nonumber
 = \frac{1}{(2\pi\hbar)^{N}} \int {\rm d}^{2N}\X ~~ W_E(\x+\X, \epsilon)~W_{E'}(\x-\X, \epsilon).
\label{Ptrans1}
\end{eqnarray}

In this way, we obtain information on the reaction of coarsegrained eigenstates to a special class of external stimuli, even if the corresponding classical dynamics is not integrable, from our knowledge of a pair spectral Wigner functions. It is important to keep in mind that
these represent a superposition of stationary density matrices. 
\footnote{This scenario is not to be confused with an initial superposition of eigenstates themselves in the same energy range. 
This would not be stationary and the decay rate would be imparted by the energy width $\epsilon$. Here each pure state in the mixture is quite stationary and the width merely expresses the possible experimental difficulty in picking out the precise initial and final eigenstates.}
The lack of normalization in the double phase space integral for $P_{EE'}(\x, \epsilon)$ follows from the focus on energy transitions: the probability of an energy transition is increased by high energy densities in comparison with the probability of a transition between the corresponding normalized microcanonical distributions.

It should be emphasized that, so far, no approximations have crept into (1.15). The Weyl propagator can be expressed as a phase space path integral \cite{Report}. This leads
to a path integral for the spectral Wigner function, which is a regularized Fourier transform of the Weyl propagator, and hence to the 
path integral for the transition density itself derived in Appendix B. 

Approximations for the spectral Wigner functions are reviewed in the following section, highlighting the important role of centre sections, that is, sections of the energy shell by their own reflection, and the mappings that trajectory segments generate on such sections. The sections shrink to zero as the reflection point $\x$ approaches the energy shell, which is a caustic of the SC spectral Wigner function. Since the only role of the energy smoothing parameter is to cut off the SC contribution of long trajectories, only the shortest trajectory segments are relevant near the caustic for sufficiently large $\epsilon$. This regime of a wide enough energy window, for a zero-width $\delta$-function on the energy shell to be a sufficiently valid spectral Wigner function, is the basis of our approximations in sections 3 and 5. It is only for $N=1$ that the broader realization of the spectral Wigner function as an Airy function is needed.

The integration for the transition (1.15) depends on a generalization of the centre section involving different energy shells, which is discussed in section 3. The  classical form of the transition probability obtained in section 4 relies on the Poisson bracket between the Hamiltonian and its reflection along this section, which is also assumed to be small here, that is, $\x$ lies close to both classically narrow shells. But then one must sort out the limit where the section can be arbitrarily small, with the Poisson bracket denominator also vanishing. Thus, using local quadratic coordinates introduced in section 3, it is found in section 5 that the simple classical approximation furnishes a transition probability that goes smoothly to zero along the caustic unless $N=1$. Of course, this case is intrinsically nonchaotic, but it is worthwhile to understand how the more refined Airy function approximation resolves the problem in section 6. The final discussion in section 7 dismisses any possible invocation of quantum ergodicity to extend these simple results to the transitions between individual eigenstates of classically chaotic systems and indicates the need of further SC approximations to refine our picture of transitions between classically narrow energy shells.

\section{Aspects of the spectral Wigner function}

The SC approximation for $W_E(\x,\epsilon)$ is based on trajectory segments in the $E$-shell with both tips centred on $\x$, that is,
\be
H(\x+\X) = H(\x-\X) = E
\ee 
\cite{Ber89,Ber89b,Report}. Since Berry presents a careful discussion of its features for different parameter regimes in \cite{Ber89},
only the relevant points for the present purpose will be developed here.
As usual with SC methods, a search is needed for trajectories that are seeded by their initial conditions, but a shortcut going back to \cite{AlmHan82} is here instrumental. First one defines for each fixed centre $\x$ the pair of Hamiltionians
\be
H_+(\X) \equiv H(\x+\X) ~~~ {\rm and} ~~~ H_-(\X) \equiv H(\x-\X) ~,
\ee
so that the origin is merely translated in $H_+(\X)$, whereas $H_-(\X)$ is its reflection. It follows  that all the symmetric pairs of points, which are possible candidates for the endpoints of the appropriate trajectories, lie in the intersection of the energy shells of $H_+(\X)$ and $H_-(\X)$.   

In the case that $N=1$, these are discrete pairs of points and the search is ended, but for higher dimensions the energy shell is $(2N-1)$-D and so the intersection is $2(N-1)$-D, a reflection symmetric section of the original energy shell called the {\it centre section} in \cite{Report}. Its geometry and the relation with more familiar Poincar\'e sections will be the subject of study in the following section, but emphasis is here on the fact that trajectories crossing the section map it back to itself. If $\x$ lies close to the energy shell, the first return is carried out by very short nearly straight trajectories, whereas further returns, though guaranteed by Poincar\'e recurrence \cite{ArnAvez,livro}, may involve very long orbits. The two crucial points here are, first, that the trajectories which contribute to the SC approximation are the fixed points of the {\it centre map} (for any number of windings).

The second point concerns the effect of the width $\epsilon$ of the energy shell, which is to exponentially dampen the contribution of long trajectories. Thus, if a periodic trajectory of period $\tau$ intersects a small centre section, it will only contribute significantly 
if $\epsilon~\tau < \hbar$. One should note that periodic trajectories that do not intersect the section give at most evanescent contributions and we may be satisfied that only the very small trajectories contribute if none of the periodic trajectories traversing the
section satisfy this condition, which still allows for a classically narrow energy shell. This restriction will be assumed here on.    

The simplest approximation, for a sufficiently small centre section, is to model each short trajectory segment by its tangent $\dot{\x}~\delta t$, furnished by Hamilton's equations. This leads to the classical result
\be
W_E(\x, \cdot) \approx \delta(H(\x)-E),
\label{clasWig}
\ee
in terms of a true Dirac $\delta$-function with no trace of $\epsilon$. The spectral Wigner function has collapsed onto the energy shell and coincides with the classical microcanonical distribution.

The inclusion of a first correction to the trajectories, allowing for a quadratic curvature, softens the spectral Wigner function into an Airy function \cite{Abramowitz}:
\be
W_E(\x, \cdot) \approx \frac{1}{\gamma(\x)}{\rm Ai}\left[ \frac{H(\x) - E}{\gamma(\x)}\right] ~.
\label{Airy}
\ee
Here
\be
\gamma(\x) = \frac{(\hbar^2 \dot{\x}~ \mH_{\x}~ \dot{\x})^{1/3}}{2} = \frac{(\hbar^2 \dot{\x}~ \wedge \ddot{\x})^{1/3}}{2},
\label{width}
\ee
with the Hessian matrix of the Hamiltonian evaluated at the reflection centre 
\be
\mH_{\x} = \frac{\der^2}{\der\x^2} H(\x)~.
\ee
The Airy function integrates as the Dirac $\delta$-function in the limit that $\gamma(\x) \rightarrow 0$. The energy shell separates the evanescent region of the Airy function outside, from the oscillatory region inside; it is the fold caustic of the SC spectral Wigner function. This recovery of the simple classical form of the spectral Wigner function follows from the standard SC assumption that Planck's constant is sufficiently small, but it also depends on the the curvature correction to the linear approximation of the Hamiltonian. It may still seem strange that the energy width 
$\epsilon$ does not entre into \eref{Airy} nor \eref{clasWig}, but the original SC deduction in \cite{Ber89b} makes it clear that its only role
is to cut off further contributions from longer orbits, which must be very close to periodic orbits. The condition for the approximation \eref{Airy} to hold is then $\hbar/\tau_1 << \epsilon << \gamma(\x)$, if the shortest orbit has period $\tau_1$.
   
As the reflection centre is removed from the neighbourhood of the energy shell, the lengthening trajectory segments that first reintersect the growing centre section need a more complex uniform approximation, which is still based on the Airy function. On the other hand, the latter can then be expressed in terms of the cosine of a phase that is proportional to the trajectory's centre action. The focus in this initial study is the caustic contribution, for which the above approximations suffice. We shall see in section 4 that even the extreme classical approximation \eref{clasWig} supplies a finite transition probability, except in the (nonchaotic) case where $N=1$. Then one must resort to the Airy function \eref{Airy} to understand the behaviour near the caustic. To this end, we first discuss the geometry of small centre sections, while allowing for a different energy for each shell.

\section{Small sections of the energy shell}

For the purpose of evaluating the energy transition integral (1.15), it is worthwhile to allow the reflected shell 
to have a different energy $E'$. Thus, the extended centre section is defined by the pair of equations
\be
H_+(\X) = H(\x+\X) = E ~~~ {\rm and} ~~~ H_-(\X) =  H(\x-\X) = E'
\label{section}
\ee
and it is only in the diagonal case where $E=E'$ that this $2(N-1)$-D surface is symmetric about $\x$.  
 
In a region surrounding a single minimum of the Hamiltonian, an energy shell has the topology of a $(2N-1)$-D sphere in the $2N$-D phase space, so that its familiar Poincar\'e section by a plane is also topologically a sphere. Its usual depiction as a disk (in the case where $N=2$) is a projection of a hemisphere onto the section plane. Just as the boundary of this disk is not generally convex, so the full section
need not be convex and the same goes for the centre section. Given the overall topology, there must be regions in which a local quadratic approximation will be reducible to a high dimensional ellipsoid, though this cannot be assumed everywhere. 

All the same, following \cite{Report} one can investigate the case where $|E-E'| << E$, $\x$ is placed close to both shells and all the eigenvalues of $\mH_{\x}$ are positive. Then the quadratic approximation to the shells are concentric ellipsoids, being that this common centre does not generally coincide with the equilibrium at the origin of the $\x$-phase space. First shifting the origin to the centre of the ellipsoids and then diagonalizing the quadratic Hamiltonian by a symplectic transformation 
\be
H(\x) \approx \frac{1}{2} {\x}~ \mH_{\x}~ \x  \mapsto \frac{1}{2} \sum_{n=1}^N \omega_n (p_n^2 + q_n^2) ~,
\label{Hpm}
\ee 
one can further simplify the Hamiltonian into spherical symmetry by scale transformations in each coordinate plane:
\be
H(\x) \approx \frac{\omega}{2} \x^2~.
\label{spherical}
\ee     
with $\omega^N = \omega_1 ... \omega_N$. 
\footnote{Care must be taken because this last step is not symplectic even though volume preserving. This does not alter the geometrical considerations which follow nor rescale the SC energy spectrum, but any dynamics, which may be examined, needs to be rescaled in the end.}   

\begin{figure}
\centering
\includegraphics[width=.6\linewidth]{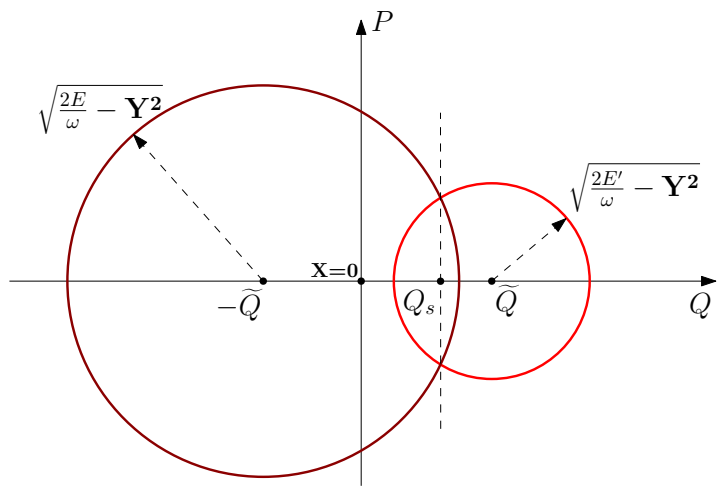}
\caption{Fixing the $2(N-1)$ transverse variables synthesized by the vector $\Y$, the pair of energy shells $H_+(\X)= E$ and $H_-(\X)= E'$ are circles with centres on the $Q$ axis at $-{\tilde Q}$ and $\tilde Q$ respectively. In this 2-D slice the section is reduced to a pair of points with $Q=Q_s$, as in the case where $N=1$.}  
\label{Fig1}
\end{figure}

Shifting again the origin to the reflection centre $\x$, one regains a pair of Hamiltonians $H_{\pm}(\X)$ as defined in \eref{section}, so as to  analyze the intersection of the spherical shells $H_+(\X)= E$ and $H_-(\X) = E'$. Since the centres ${\tilde\X}_{\pm}$ of both these spheres are collinear with the origin at the reflection centre $\X=0$, one can follow \cite{Report} in choosing a symplectic coordinate system such that this line becomes the $X_{2N}$ axis, that is, 
\be
\tilde\X = (\Y=0, X_{2N-1}=0, X_{2N}=\tilde Q),
\ee 
where $\Y$ is a generic point in the remaining $2(N-1)$-D plane. Calling $P=X_{2N-1}$ the coordinate conjugate to $X_{2N}=Q$,
our working coordinates become $(\Y, P, Q)$, which may also be employed for the original $\x$ variables, except for the change of origin to $-\tilde Q$, the centre of the unevolved sphere. Thus, $\tilde Q$ is the single scalar parameter 
defining the reflections. The equations for the centre section are then spelled out as
\be
\fl \frac{\omega}{2}~[(Q+\tilde Q)^2+P^2+\Y^2]= E ~~~{\rm and} ~~~ \frac{\omega}{2}~[(Q-\tilde Q)^2+P^2+\Y^2]= E'~,
\label{Hpm2}
\ee
which, subtracted, furnish the linear equation in $Q$:
\be
(Q+\tilde Q)^2 - (Q-\tilde Q)^2 = 4 {\tilde Q}~Q = \frac{2}{\omega} (E - E').
\ee
Hence, the centre section coincides in this simple case with the Poincar\'e section by the plane with
\be
Q_s = \frac{1}{2\omega \tilde Q} (E-E').
\ee    
In the diagonal case studied in \cite{Report} there is complete symmetry about the reflection centre at the $\X$-origin, so that
$Q_s = 0$. In general it will be assumed that
\be
\sigma = 2 \frac{E-E'}{E+E'} = \frac{E-E'}{\bar E}
\ee
is a small parameter, so that the section coordinate
\be
Q_s = \frac{\bar E}{2\omega \tilde Q}~ \sigma
\ee
lies close to the origin.
Fixing a value of $\Y^2$, we obtain a slice through the pair of energy shells as a pair of circles in the $(P,Q)$ plane, as shown in the Fig. 1. The value of $P(\Y^2)$, the momentum conjugate to $Q_s$ at the section, can be obtained from either equation in \eref{Hpm2} or, more symmetrically, from  their sum
\be
Q_s^2+{\tilde Q}^2+ P^2+ \Y^2 = \frac{1}{\omega}(E + E') ~.
\ee
Then, one may define 
\be
\Y_M^2 = \frac{1}{\omega}(E + E') - (Q_s^2 + {\tilde Q}^2) = P_s^2,
\ee
that is, $\Y_M$ is radius of the Poincar\'e section in the $\Y$ space, so that
\be
P(\Y^2) = \pm \sqrt{\Y_M^2 - \Y^2}
\ee 
and $P_s$ is the momentum at the section for $\Y=0$.
The locus where the section shrinks to zero, i.e. $\Y_M=P_s=0$ in the spherical example, is the caustic where the two shells are tangent. 
If $E=E'$, this is just $H(\x)= E$, since the tangent plane to the shell is invariant with respect to a reflection through one of its points. 
The off-diagonal case was discussed in \cite{Oz84} for $N=1$. Its generalization is a surface lying between $E$ and $E'$, which may be approximated by the shell with the average energy, in the limit $E' \rightarrow E$.

\section{Classical approximation of the transition probability density}

Assuming that the width of the energy shells has been chosen such that $\epsilon~\tau_1 >>\hbar$, where $\tau_1$ is the smallest period of a periodic trajectory, so that only the shortest trajectory segments need be considered and recalling that the Airy function integrates as a $\delta$-function, we obtain the simplest approximation for the transition probability density by inserting \eref{clasWig} into (1.15):
\be
P_{EE'}(\x,\cdot) \approx  \frac{1}{(2\pi\hbar)^{N}} \int {\rm d}^{2N}\X ~ \delta(H_+(\X) - E) ~ \delta(H_-(\X) - E') ~,
\label{Ptrans2}
\ee
where it should be recalled that the implicit dependence on the reflection centre $\x$ lies in the definition \eref{section} of the displaced Hamiltonians $H_\pm$.
Let us first start with $N=1$, so that the closed energy curves intersect at a pair of points $\X_1$ and $\X_2$ and 
\begin{eqnarray}
\fl P_{EE'}(\x,\cdot) \approx  \frac{1}{2\pi\hbar}\sum_{j=1}^2 \int dH_+ dH_- \left| \frac{\der(H_+ H_-)}{\der(P,Q)} \right|_{\X_j}^{-1}
                           \delta(H_+ - E) ~ \delta(H_- - E') \\ \nonumber
                   =  \frac{1}{2\pi\hbar}\sum_j |\{H_+,H_-\}_{\X_j}|^{-1} ~,
\label{Ptrans3}
\end{eqnarray}
where the Poisson bracket of the pair of Hamiltonians \cite{Arnold}
\be
\{H_+,H_-\} \equiv \frac{\der H_+}{\der P}~ \frac{\der H_-}{\der Q} - \frac{\der H_+}{\der Q} ~ \frac{\der H_-}{\der P}
\ee
coincides with the Jacobian determinant of the integrand. It goes to zero at the caustic tangency, so that the classical singularity can only be resolved by the Airy function approximation, as will be shown in section 6.

For higher $N$ there is an excess of integration variables in \eref{Ptrans2} over the arguments of the $\delta$-functions, but one can single out the 2-D {\it velocity plane} spanned by the pair of Hamiltonian velocity vectors 
\be
{\dot \X}_{\pm}(\X) = \J \frac{\der H_{\pm}}{\der \X}(\X) ~,
\label{velocities}
\ee
where the standard symplectic matrix in Hamilton's equations is
\be
\J = \left(
\begin{array}{cc}
     0 & -1 \\
     1 & 0 
\end{array}
\right) 
\ee 
in terms of $(\mathbf{P},\mathbf{Q})$ blocks at any point $\X_s$ on the centre section \eref{section}. Thus, the tangent plane
to each shell is determined by 
\be
{\dot \X}_+ \wedge (\X-\X_s)=0 ~~~ {\rm and} ~~~  {\dot \X}_- \wedge (\X-\X_s)= 0 ~,
\label{tangent}
\ee
whereas their intersection is the tangent plane to the section. 
\footnote{Recall that the skew product is defined as $\x \wedge \y \equiv (\J\x)\cdot \y$.}
Furthermore, this pair of equations guarantees that all vectors on the tangent plane are skew orthogonal to the velocity plane. This provides the freedom to construct a symplectic coordinate system, such that
$\X = (\Y, P, Q)$, just as in the simple spherical example, with the components of $\Y$ lying in the tangent plane and $\Y=0$ when $\X=\X_s$, whereas $(P,Q)$ lies in the velocity plane. Clearly, the Hamiltonians do not depend locally on $\Y$, so one can integrate, just as for $N=1$, over $P$ and $Q$ to obtain 
\be
\int dP ~dQ ~ \delta(H_+(\X) - E) ~ \delta(H_-(\X) - E') =  |\{H_+,H_-\}_{\X_s}|^{-1} ~.
\ee    
Then the integral over the local tangential variables around the the $2(N-1)$-D surface of the centre section 
supplies the approximate transition probability density
\be
P_{EE'}(\x,\cdot) \approx  \frac{1}{(2\pi\hbar)^{N}}\oint_{\X_s} d^{2(N-1)}\X ~|\{H_+,H_-\}_{\X_s}|^{-1} ~.
\label{Ptrans4}
\ee

This is the pure classical result, because \eref{Ptrans2} can be interpreted as the joint probability of obtaining the energy $E'$ after a classical phase space reflection, given the original microcanonical probability distribution with energy $E$. No constraint to a small section is assumed so far, that is, generally the centre section cannot be approximated by an ellipsoid and the integration should be numerical. 
Of course, the Dirac $\delta$-function is not a proper probability distribution, which induces the inappropriate singularity of the transition probability density along the caustic for $N=1$. The correct softening of this distribution is in terms of the Airy function for quantum mechanics, but this is not positive, so some alternative softening is also necessary in the classical context. In the next section it is shown that the integral is well behaved in all other cases, which include chaotic contexts.

\section{Local spherical approximation near the caustic}

It was argued in section 3 that a local spherical approximation encapsulated the essential features of both the fixed $E$-shell and the $E'$-shell reflected through a centre near both these shells. Then, for the pair of Hamiltonians $H_{\pm}(\X)$, given by \eref{section} within the spherical approximation \eref{spherical}, the phase space velocities assume the coordinates
\be
{\dot X}_{\pm} = \omega~ \J(\x\pm\X) =\omega~(\J\Y, -(Q_s \pm \tilde Q), P(\Y^2))~ ,
\ee             
where the symplectic matrix $\J$ in the second equality is restricted to the $2(N-1)$-D $\Y$-subspace. It follows that, even though the velocity plane does not coincide with the plane $\Y=0$, both velocities have the same component in this subspace, so that the Poisson bracket 
\be
{\dot X}_+ \wedge {\dot X}_- = \{H_+,H_-\}(\Y^2) = 2\omega^2 {\tilde Q}~ P(\Y^2)= \pm 2\omega^2 {\tilde Q} \sqrt{\Y_M^2 - \Y^2}
\ee
depends only on the velocity components in the $(P,Q)$ plane.

Thus, the integral \eref{Ptrans4} over the spherical section becomes an integral over the $2(N-1)$ sphere with radius $|\Y_M|$ 
and both hemispheres labled by the $\pm$ signs give equal contributions to the transition probability density
\be
P_{EE'}(\x,\cdot) \approx \frac{1}{(2\pi\hbar)^{N}}  \int_{\Y^2 <\Y_M^2} ~ \frac{d\Y}{\omega^2 {\tilde Q}\sqrt{\Y_M^2 - \Y^2} } ~,
\label{Ptrans5}
\ee
with a singularity of the integrand along the border of the disk even for a finite section. But reverting to spherical coordinates in the $2(N-1)$-D $\Y$-space and recalling that the $n$-D area of a sphere with radius $|\Y|$ is $C_n |\Y|^n$, where
\be
C_n = \frac{2\pi^{(n+1)/2}}{\Gamma(n+1)/2} ~,
\ee
reduces this to the single integral
\be
P_{EE'}(\x,\cdot) \approx  \frac{1}{(2\pi\hbar)^{N}} \int_0^{|\Y_M|} ~ \frac{C_{2N-3} |\Y|^{2N-3}~d\Y}{\omega^2 {\tilde Q}\sqrt{\Y_M^2 - \Y^2} } ~,
\label{Ptrans6}
\ee
which can be successively integrated by parts for any $N>1$.

For $N=2$ 
\be
\fl P_{EE'}(\x,\cdot) \approx \frac{ |\Y_M|}{2\pi(\hbar\omega)^2 {\tilde Q}}
= \frac{1}{2\pi(\hbar\omega)^2 {\tilde Q}} ~\sqrt{\frac{E+E'}{\omega}+ \left(\frac{E-E'}{2\omega \tilde Q}\right)^2 - {\tilde Q}^2}.
\label{Ptrans7}
\ee
Therefore, not only does the classical integral for the transition probability density converge, 
in spite of the singularity at the border of the disk,
but the shrinking of the section at the caustic dominates the singularity. Even though it is the second expression that provides the transition probability density in terms of the given variables, the important geometrical parameter is seen to be the radius $|\Y_M|$ of the section disk. For N=3, one obtains
\be
P_{EE'}(\x,\cdot) \approx \frac{1}{2\pi\hbar^{3}} \frac{1}{\omega^2 {\tilde Q}} |\Y_M|^3
\label{Ptrans8}
\ee
and in general $P_{EE'}(\x,\cdot) \propto |\Y_M|^{2N-3}$. 

Beyond our simple spherical model, one can still associate this radius to the linear scale of the $2(N-1)$-D centre section. 
The transition probability density in the classical approximation \eref{Ptrans4}, an integral over a general centre section,
can then be performed numerically without any qualms as to its convergence, no matter how close the reflection point may be to the caustic.

\section{Resolution of the singularity for a single degree of freedom}

In contrast to the extreme classical limit of the transition probability density based on the sharp approximation of the spectral
Wigner functions by Dirac $\delta$-functions, their improved Airy function approximation \eref{Airy} leads to an integrand in (1.15),
which is not evanescent only within the intersection of the phase space volumes of the pair of energy shells. It is the interference of the
oscillatory fringes of both Airy functions (depicted for $N=1$ in \cite{SarOA16}) that allows for the predominance of the section in the simple
classical approximation. In any case, the dominant oscillatory region will be classically small near a caustic, which still allows for
many interfering fringes in the SC limit. 

The only case where the classical approximation for the transition probability density is singular at the caustic occurs if $N=1$. This is not relevant for the study of quantum chaos, but it is a first indication of how the improved Airy function approximation \eref{Airy} becomes essential.
Restricting again consideration to the spherical shells, which are here merely circles, their intersection reduces to the pair of points in Fig. 1 with $Q=Q_s$ and $\Y=0$. In these coordinates, the single scalar parameter specifying the reflection centre $\x$ 
is $\tilde Q$, that is, $(-\tilde Q,P=0)$ is the origin of the original coordinate system.
Then given the expressions \eref{Hpm} for $H_{\pm}(\X)$ in this special case, that is,
\be
H_{\pm}= \frac{\omega}{2}~\left[(Q\pm\tilde Q)^2+P^2 \right] 
\ee
 the widths \eref{width} of the corresponding Airy functions are explicitly
\be
\gamma_{\pm} = \frac{\omega \hbar^{2/3}}{2} ~ \left[(Q\pm\tilde Q)^2 + P^2 \right]^{1/3} 
\label{gpm}
\label{widthsa}
\ee
and the integral for the transition probability density becomes
\be
P_{EE'}(\tilde Q,\cdot) \approx \int\frac{dP}{2\pi\hbar}  \int \frac{dQ}{\gamma_+ \gamma_- } 
{\rm Ai}\left[ \frac{H_+ - E}{\gamma_+}\right]~
{\rm Ai}\left[ \frac{H_- - E'}{\gamma_-}\right]~.
\label{Ptrans9}
\ee

In the regime that is here focused, the arguments of both Airy functions are small in the neighbourhood of both energy shells,
but neither of the energies are assumed to be small and our parameter $\tilde Q$ is also of the order of the radius of the circles.
It is then convenient to reparametrize the arguments of the Airy functions in terms of the positions
\be
Q_+ = -{\tilde Q} + \sqrt{\frac{2}{\omega} E} ~~~~~~~ Q_- = {\tilde Q} - \sqrt{\frac{2}{\omega} E'} ~,
\ee
where the energy shells of $H_\pm$ intersect the $Q$-axis. It is only within the region $Q_- \leq Q \leq Q_+$ (which contains $Q_s$) 
and $P^2 \leq P_s^2$ that the product of Airy functions is oscillatory, so that the contribution to the integral outside this range is negligible. 
The caustic condition is just $Q_+=Q_-$, for which the circles touch and $P_s=0$, that is
\be
\tilde Q = Q_c \equiv \frac{1}{2 \omega} \left[\sqrt{E} + \sqrt{E'}~  \right] ~
\ee
and the distance from the caustic is just
\footnote{The condition for overlap of the spheres, including oscillations for the product of Airy functions, is $\tilde{Q}< Q_c$.
One should distinguish $Q_c$, which is a particular value of $\tilde Q$ depending on the pair of energies, 
from the section parameter $Q_s$ in section 3, which depends on the choice of $\tilde Q$, as well as the pair of energies.}  
\be
\tilde Q -Q_c = \frac{1}{2} (Q_{-} - Q_{+}) ~.
\ee
In the diagonal case, $E=E'$, $Q_-=-Q_+$, so that $Q_c=0$.
In any case, as discussed in section 3, one can assume in the parameter regime close to the caustic 
that both $|Q_\pm| << \tilde Q$, entailing $|Q_s| << \tilde Q$.

Thus, the integral is dominated by the region where $|Q| << \tilde Q$ and $P$ is small,
so that all the terms are small in 
\be
\fl H_+ - E= \frac{\omega}{2}~ \left[P^2 + Q^2 +2\tilde{Q}Q - 2\tilde{Q} Q_+ - Q_+^2 \right]
\approx \frac{\omega}{2}~ \left[P^2 +2\tilde{Q}Q - 2\tilde{Q} Q_+ - Q_+^2 \right]
\ee
and
\be
\fl H_- - E'= \frac{\omega}{2}~ \left[P^2 + Q^2 -2\tilde{Q}Q + 2\tilde{Q} Q_- - Q_-^2 \right] 
\approx \frac{\omega}{2}~\left[P^2 -2\tilde{Q}Q + 2\tilde{Q} Q_- - Q_-^2 \right] .
\ee
In contrast, the widths \eref{widthsa} are not small;
their value near the respective energy shells are dominated by $\tilde Q^2$ and so both widths can be approximated
by their value at the caustic:
\be
 \gamma_{\pm}\approx \gamma_0 =  \frac{\omega}{2} (\hbar ~ \tilde Q)^{2/3} ~.
\ee
Then the integral assumes the explicit form
\begin{eqnarray}
\fl P_{EE'}(\tilde Q,\cdot) \approx \int \frac{dP}{2\pi \hbar} \int \frac{dQ}{\gamma_0^2} \\  \nonumber
{\rm Ai}\left[   \frac{P^2 +2\tilde{Q}Q - 2\tilde{Q} Q_+ - Q_+^2}{(\hbar\tilde Q)^{2/3}}\right]~  
{\rm Ai}\left[  \frac{P^2 -2\tilde{Q}Q + 2\tilde{Q} Q_- - Q_-^2}{(\hbar\tilde Q)^{2/3}}  \right]~.
\label{Ptrans10}
\end{eqnarray}

At this point, one invokes the identity \cite{Abramochkin}
\be
\int dq ~ {\rm Ai}(aq+b) ~ {\rm Ai}(-aq+c) = \frac{1}{2^{1/3}a} {\rm Ai}\left[\frac{1}{2^{1/3}}(c+b)\right]~,
\ee 
so that, choosing the constant $a= 2({\tilde Q}/\hbar^2)^{1/3} = \omega \tilde{Q}/\gamma_0$ and reverting back to the energy parameters, one then obtains
\be
\fl P_{EE'}(\tilde Q,\cdot) \approx  \frac{1}{2\pi\hbar{\tilde Q}\omega}\int dP ~ 
\frac{1}{ 2^{1/3}\gamma_0} ~ {\rm Ai}\left[ \frac{1}{2^{1/3}\gamma_0}~[\omega({\tilde Q}^2+P^2) - (E+E')]\right]
\label{Ptrans10} ~.
\ee
Thus, the transition density reduces to a single integral over an Airy function. Moreover, one recognizes that $2^{1/3}\gamma_0$
is actually the width of the SC approximation of a spectral Wigner function, namely
\be
W_{E+E'}(\X) \equiv \frac{1}{2^{1/3}\gamma_0} ~ {\rm Ai}\left[ \frac{1}{2^{1/3}\gamma_0}~[\omega({\tilde Q}^2+P^2) - (E+E')]\right] ~,
\ee
for the $E+E'$ energy shell of a harmonic oscillator with double the frequency. The remaining integral is identical to the projection
of this spectral Wigner function onto the position coordinates to obtain the probability density near the caustic, which here
lies at $\omega {\tilde Q}^2 = E+E'$ or equivalently $(\omega/2) {\tilde Q}^2 =\bar E$. This projection was performed by Berry 
in \cite{Ber89b} for $N \geq 1$, neglecting
the dependence of the width on $P^2$, due to the proximity of the caustic. Here the potential is merely $V(\tilde Q)  = \omega {\tilde Q}^2$,
so that one obtains 
\be
P_{EE'}(\tilde Q,\cdot) \approx \frac{1}{(\omega \hbar) \sqrt{\omega{\tilde Q}^2 ~ 2\gamma(0)}}
\left\{{\rm Ai}\left[\frac{ \omega{\tilde Q}^2- (E+E')}{2\gamma(0)}\right]\right\}^2 ~,
\label{Ptrans11}
\ee  
where both factors in the denominator of the amplitude of the transition probability density have the required energy dimension.
Seeing as $\frac{\omega}{2}{\tilde Q}^2=H(\x)$ in the original variables, the argument of the Airy function has the same form as those of the spectral Wigner functions in the integrand for the transition probability, but with their average energy.
In the diagonal case $E=E'$, the non-transition density is merely the square of the spectral Wigner function, in close analogy to the
permanence probability of a pure state in the exact formula \eref{Pkk}.

The presence of Planck's constant in the width $\gamma_0$ implies that there is a true peak of the transition density at the caustic,
where the simple classical approximation becomes singular. For higher dimensions, that is, $N>1$, it was not necessary to invoke the full
Airy function approximation of the spectral Wigner functions, being that there is no classical singularity. Even so, it may be worthwhile
in the future to attempt to derive full SC approximations in these well behaved cases, in contrast to the classical results. One should note
that in the projection integrals in \cite{Ber89b}, the analogous result was that only in the $N=1$ case did the projection show sign of a caustic.

We have generally suppressed any indication of the width parameter $\epsilon$ for the energy shells, assuming that it is sufficiently large to leave only the contribution of the shortest trajectory in the SC spectral Wigner function. In the present case of a single degree of freedom, one can investigate the opposite regime where $\epsilon \rightarrow 0$, in which one picks out the transition Wigner function for a pair of pure eigenstates studied in \cite{Oz84}. There the expressions are in terms of the action variable for the closed energy curves instead of the Hamiltonian, but one can consider simply that $\omega=1$. The main difference is that the transition Wigner function then has a new term for each return of the trajectory to either of the two points in the section, as well as the short segment that has been studied here. 
\footnote{The simplicity of the final result follows from the Bohr quantization of the action variable for an eigenstate.}
The explicit form of this term of the transition Wigner function in the limit very close to the caustic was not presented in \cite{Oz84}, but its diagonal form (no transition) is just the SC Wigner function \cite{Ber77}.

\section{Discussion: quantum ergodicity?}

Notwithstanding our careful use of the finite width of the energy shell in the employment of the spectral Wigner function, a coarsegrained representation of a microcanonical (non-normalized) mixed state, quantum ergodicity does hold in a weak sense for systems with classical chaotic motion. Given the nontrivial constructions that are needed to access SC approximations for chaotic eigenstates \cite{Bog90,BerKea}, it is remarkable that such a simplification can be proven \cite{Shnirelman} for some hard chaotic systems. 
In a nutshell, expected values in a single pure state of smooth observables have high probability of being close to that derived from the classical microcanonical distribution. The integral for the expectation \eref{average} in the Wigner-Weyl representation has the same form as the integral 
(1.15) for the transition probability after a reflection, so it may seem that quantum ergodicity could also be invoked here in certain contexts. 

Quantum ergodicity \cite{Shnirelman} does not presuppose that the Wigner function for a pure state is nonoscillatory, but short wave oscillations should not be a feature of the test function in the expectation integral. The problem is that the Wigner function is never exempt from such oscillations. Not only are they present in the SC approximation for $N=1$, but they must be present for all the highly excited eigenstates whatever the number of degrees of freedom. An intuitive argument for this was presented by Zurek \cite{Zurek}, but the clinch comes from the Fourier identity \cite{ChouVou,AlmValSar} for the correlation
\be
\fl C(\Vxi) \equiv (2\pi\hbar)^N \int d\x ~ W(\x) ~ W(\x + \Vxi)  
= (2\pi\hbar)^{-N} \int d\Veta ~ \exp\left(\frac{i}{\hbar} \Vxi \wedge \Veta \right) ~ C(\Veta)~.
\ee
This implies that large scale features of a pure state Wigner function, such as the diameter $D$ of the $E$-shell, which affect the correlation,  are accompanied by ripples of wave length $\hbar/D$ in the correlation and hence in the Wigner function itself. But this correlation integral is even closer to that for the transition probability than an expectation. Hence, it is at least hazardous to suppose that the latter can be exempt from such nonclassical ripples, with respect to changes of the reflection centre. 

Oscillations in the transition probability between pure eigenstates are progressively revealed by shrinking the energy width, as longer trajectory segments contribute to the pair of SC spectral Wigner functions. Thus, the next step will be to include their centre actions so as to evaluate higher contributions to (1.15) by stationary phase. For small sections near the caustic, the segments can be considered to be perturbations of the various windings of the shortest periodic trajectory which crosses the section, so that it has a role similar to the energy curve if $N=1$. The centre section grows as the reflection centre is removed from the energy shell, so that more short periodic trajectories traverse it. In any case, periodic trajectories that do not cross a given centre section contribute at best an evanescent term to the spectral Wigner function, whatever the width $\epsilon$. The program of orbit resummation to obtain a pure state Wigner function or transition probability distribution, such as \cite{AgFish}, generally equates the cutoff period $\hbar/\epsilon$ to half the Heisenberg time, but here too one should discriminate the periodic trajectories that contribute according to the reflection centre and its section.  

So far, it is the centre section itself for a pair of coarse-grained energy shells that furnishes the classical average of the transition probabilities of the eigenstates in classically narrow energy ranges. It is important to distinguish the $(2N-1)$-D thinness of the pair of energy shells from possibly large $2(N-1)$-D volume of their intersection. The emphasis on small centre sections described by local coordinates allows us to resolve an apparent singularity near the caustic, but the transition probability density may in principle be integrated numerically 
in \eref{Ptrans4} over large sections.

No distinction is made at this level about the dynamical structure of the trajectories in the shell. This may have regions of negative curvature 
$(N=2)$ and the local nature of the centre section in such regions is still to be investigated. 
The special nature of the unitary reflection operator that provokes these transitions, in its dual role as the operator basis for the spectral Wigner function itself, has been essential for these results. Allowing for smaller energy widths which depend on the arrangements of trajectory segments, the flexibility of the reflection operator should become a unique probe into the quantum dynamical behaviour of classically chaotic systems.

\appendix

\section{More on the phase space translation operators and the Wigner function}

The phase space reflection operators $\hat{R}_\x$ are the Fourier transforms of the translation operators
(corresponding to a phase space translations by $2\Vxi$)  \cite{Report}
\be
{\hat T}_\Vxi = \exp\left[\frac{2i}{\hbar}\Vxi \wedge {\hat \x}\right] 
= \int {\rm d}^{N} \vecq ~|\vecq+ \Vxi_{\vecq}\rangle\langle \vecq- \Vxi_{\vecq}|~~ \exp\left[\frac{2i}{\hbar}\Vxi_\vecp \cdot {\vecq}\right]  ~,
\ee 
with the similar expression
\be
\fl \hat{R}_\x = \int \frac{{\rm d}^{2N} \Vxi}{(2\pi\hbar)^N} ~ {\hat T}_\Vxi  ~  \exp\left[\frac{2i}{\hbar}\x \wedge {\Vxi}\right] 
= \int {\rm d}^{N} \Vxi_{\vecq} ~|\vecq+ \Vxi_{\vecq}\rangle\langle \vecq- \Vxi_{\vecq}| ~
\exp\left[-\frac{2i}{\hbar}\Vxi_\vecp \cdot {\vecq}\right]  ~.
\ee  
Inserted into \eref{average}, this leads to the representation of an operator $\hat O$ as
\be
O(\x) = {\rm tr} ~ \hat{R}_\x ~ \hat O = \int \frac{{\rm d}^{N} \Vxi_{\vecq}}{(2\pi\hbar)^N} ~ \langle \vecq- \Vxi_{\vecq}|\hat{O}|\vecq+ \Vxi_{\vecq}\rangle ~~ 
\exp\left[-\frac{2i}{\hbar}\Vxi_\vecp \cdot {\vecq}\right] ~,
\ee
the familiar Weyl-Wigner transform of its position matrix elements. In the case of the density operator, one obtains the Wigner function 
\eref{Wigfn}, within a normalization factor.    

Just as a general classical reflection, $\x'\mapsto -\x'+2\x$, through the point $\x$ can be considered to be a similarity transformation of the reflection through the origin, brought about by its translation to $2\x$ and back again, so is the phase space reflection operator obtained from $\hat{R}_0$ (also known as the parity operator) as $\hat{R}_\x = {\hat T}_{-\x}\hat{R}_0{\hat T}_\x$.

\section{Path integrals in phase space}

The {\it Weyl propagator} $V(\x,t)$
\be
\hat{V}(t) = \e^{-it{\hat H}/ \hbar}~,
\label{intev}
\ee
represents the unitary evolution operator generated by $\hat H$,
that is,
\be
\lim_{t\rightarrow 0} V(\x,t) = \e^{-it{H(\x)}/ \hbar}~,
\ee
where strictly $H(\x)$ is the Weyl representation of the Hamiltonian, which is semiclassically close to the classical Hamiltonian.
Then, decomposing the finite time evolution into a product of an even number $K$ of small time evolutions,
the phase space path integral for the Weyl propagator takes the form \cite{Oz92,Report}
\be
\fl V(\x,t) = \lim_{K\rightarrow \infty} \int \frac{d^{2N}\x_1~ ... ~d^{2N}\x_K}{(\pi\hbar)^{KN}}~
\exp \left[ \frac{i}{\hbar}\Delta_{K+1}(\x, \x_1, ..., \x_K)- \frac{t}{K} \sum^K_{k=1} H(\x_k) \right] ~.
\ee 
Here $\Delta_{K+1}$ is the symplectic area of the odd-sided polygon in phase space, defined uniquely by the centres of its sides 
$\x, \x_1, ..., \x_K$ \cite{Report}. These are generally very jagged polygons and it is only the special form of the 
classical variational principle derived in \cite{Oz90} that picks out approximations with small sides that approximate a single
isoenergetic trajectory in the SC approximation. Indeed, each side of the polygon centred on $\x_{k}$ is the vector
\be
\Vxi_k = -\J \frac{\der \Delta_{K+1}}{\der \x_k}(\x, \x_1, ..., \x_K) ~,
\ee
which does does not depend on its own centre $\x_k$, i. e. it is a function of all the other centres and, if one singles out $\Vxi$ centred on $\x$
as the sum of all the other sides forming an open even sided polygon, it is verified \cite{Report} that
\be
\Vxi = 2 \sum_{k=1}^K (-1)^k  ~\x_k.  
\label{side}
\ee
The stationary phase condition for integration over each centre is then
\be
\Vxi_k = \frac{t}{K} ~ \J\frac{\der H}{\der \x}(\x_k) = \frac{t}{K} ~ {\dot\x}_k ~, 
\ee 
so that the polygonal side with a stationary centre is tangent to its trajectory.

The Fourier integral over time now furnishes the path integral for the spectral Wigner function
\begin{eqnarray}
 \fl W_E(\x,\epsilon) = {\rm Re} \int_0^\infty \frac{{\rm d}t}{\pi\hbar}~\exp\left[ \frac{it}{\hbar}(E+i\epsilon)\right]~ V(\x,t) \\ \nonumber
=  \lim_{K\rightarrow \infty}{\rm Re} \int d\x_1~ ... ~d\x_K ~
\exp \left[ \frac{i}{\hbar}\Delta_{K+1}(\x, \x_1, ..., \x_K) \right] ~
\delta_\epsilon \left(E - \frac{1}{K} \sum^K_{k=1} H(\x_k) \right) ,
\end{eqnarray}
where hereon the differential volume elements for each centre area are abbreviated as $d^{2N}\x_k = d\x_k$.
(This is analogous to the derivation of the Green function from the position propagator $\langle\vecq_-|\hat{V}(t)|\vecq_+\rangle$
and indeed the spectral Wigner function is the Weyl-Wigner transform of Green function.)
So it is only the average energy of the centres of the sides of the polygonal open path that is constrained to be near the given energy $E$,
with no restriction on their energy dispersion.

Finally the path integrals for the spectral Wigner function of energy $E$ and its reflected pair of energy $E'$ can both be inserted into (1.15).
Defining the centres $\x_k=\x +\X_k$ and $\x_{k'}=\x +\X_{k'}$, this becomes  
\begin{eqnarray}
\fl  P_{EE'}(\x, \epsilon)  = \frac{1}{(2\pi\hbar)^{N}} \int {\rm d}\X ~ W_E(\x+\X, \epsilon)~W_{E'}(\x-\X, \epsilon) \\   \nonumber
= \lim_{K,K'\rightarrow \infty}{\rm Re} \int d\X_1~ ... ~d\X_K d\X_{1'}~ ... ~d\X_{K'}   \\  \nonumber
\delta_\epsilon \left(E - \frac{1}{K} \sum^K_{k=1} H(\x+\X_k) \right) ~ 
\delta_\epsilon \left(E' - \frac{1}{K'} \sum^{K'}_{k'=1} H(\x+\X_{k'}) \right) 
 \\  \nonumber
\frac{1}{(2\pi\hbar)^{N}} \int {\rm d}\X ~ 
\exp \left[ \frac{i}{\hbar}\{\Delta_{K+1}(\X, \X_1, ..., \X_K) + \Delta_{K'+1}(-\X, \X_{1'}, ..., \X_{K'}) \}\right] ~,
\end{eqnarray}
using the invariance of the polygonal action with respect to the origin of the centres of its sides. This is also preserved by a
full reflection, that is, $\Delta_{K+1}(-\x, -\x_1,...,-\x_{K} = \Delta_{K+1}(\x, \x_1,...,\x_{K})$, so that reversing the sign
of all the $\X_{k'}$ integration variables leads to the alternative form
\begin{eqnarray}
\fl  P_{EE'}(\x, \epsilon) = \lim_{K,K'\rightarrow \infty}{\rm Re} \int d\X_1~ ... ~d\X_K d\X_{1'}~ ... ~d\X_{K'}   \\  \nonumber
\delta_\epsilon \left(E - \frac{1}{K} \sum^K_{k=1} H(\x+\X_k) \right) ~ 
\delta_\epsilon \left(E' - \frac{1}{K'} \sum^{K'}_{k'=1} H(\x-\X_{k'}) \right) 
 \\  \nonumber
\frac{1}{(2\pi\hbar)^{N}} \int {\rm d}\X ~ 
\exp \left[ \frac{i}{\hbar}\{\Delta_{K+1}(\X, \X_1, ..., \X_K) + \Delta_{K'+1}(\X, \X_{1'}, ..., \X_{K'}) \}\right] ~,
\end{eqnarray}
where the pair of polygons share a the common centre $\X$, whereas the other centres satisfy either the constraints defined by the Hamiltonian,
or its reflection through $\x$.

The dependence of $\Delta_{K+1}(\X, \X_1, ..., \X_K)$ on $\X$ is just $\Vxi\wedge \X$, so that the integral over $\X$ constrains
the pair of open polygonal paths with side $\Vxi$ and $\Vxi'$ centred on $\X$ to close on each other, so as to form
a single closed polygon with $K+K'$ sides and the total symplectic area
\be
\fl \Delta_{K+K'}(\X_1, ..., \X_K,\X_{1'}, ..., \X_{K'}) = \Delta_{K+1}(\X, \X_1, ..., \X_K) + \Delta_{K'+1}(\X, \X_{1'}, ..., \X_{K'}).
\ee
Thus the path integral for the energy transition density becomes
\begin{eqnarray}
\fl  P_{EE'}(\x, \epsilon)  = \lim_{K,K'\rightarrow \infty} \int d\X_1~ ... ~d\X_K d\X_{1'}~ ... ~d\X_{K'}  ~
\delta(\Vxi-\Vxi') \\  \nonumber
\delta_\epsilon \left(E - \frac{1}{K} \sum^K_{k=1} H(\x+\X_k) \right) ~ 
\delta_\epsilon \left(E' - \frac{1}{K'} \sum^{K'}_{k'=1} H(\x-\X_{k'}) \right) 
 \\  \nonumber
\exp \left[ \frac{i}{\hbar}\{ \Delta_{K+K'}(\X_1, ..., \X_K,\X_{1'}, ..., \X_{K'}) \}\right] ~.
\end{eqnarray}

It should be noted that $\Delta_{K+K'}$ is an even sided polygon, which is not uniquely defined by the centre of its sides, nor are these constraint free \cite{Report}. Indeed, arbitrary centres $\x_k$ and $\x_{k'}$ support an open polygonal line, which is closed by the condition
\be
\sum_{k=1}^K (-1)^k  ~\x_k + \sum_{k'=1}^{K'} (-1)^{k'}  ~\x_{k'} =0 ~,
\ee    
according to \eref{side}. Therefore, the shorthand for the argument of the Dirac delta function over chords in (B.11) is explicitly
\be
\delta(\Vxi-\Vxi') = \delta \left(\sum_{k=1}^K (-1)^k  ~\X_k + \sum_{k'=1}^{K'} (-1)^{k'}  ~\X_{k'}\right) ~.
\ee   
It is shown in \cite{Report} that closed even sided polygons with such given centres are not unique, but they all have the same symplectic area.

\section*{Acknowledgments}
I thank Gabriel Lando for stimulating discussions and his help in preparing the manuscript.
Both referees stimulated important additions and essential clarifications of the original manuscript.
Partial financial support from the National Institute for Science and Technology - Quantum Information
and CNPq (Brazilian agencies) is gratefully acknowledged.

\end{document}